---------------------------------------------------------------------------
############################## LaTex Copy #################################
---------------------------------------------------------------------------

\documentstyle[12pt]{article}

\begin{document}

\begin{center}

{\Large \bf Duffin-Kemmer-Petiau equation\\
on the quaternion field}

\vspace*{3cm}

Stefano DE LEO$^{\; a)}$

\vspace*{1cm}

{\it Universit\`a  di Lecce, Dipartimento di Fisica\\
Istituto di Fisica Nucleare, Sezione di Lecce\\
Lecce, 73100, ITALY}

\end{center}

\pagebreak

\begin{abstract}

We show that the
Klein-Gordon equation on the quaternion field is equivalent to a pair of
DKP equations. We shall also prove that this pair of DKP equations
can be taken back to a pair of {\em new} KG equations. We shall emphasize the
important difference between the standard and the {\em new} KG equations.
We also present some qualitative arguments, concerning the possibility of
interpreting {\em anomalous} so\-lution, within a quaternion quantum field
theory.

\end{abstract}

\vspace*{1cm}
\noindent{\footnotesize a) e-mail: {\em Deleos@Le.Infn.It}}

\pagebreak

\section{Introduction}

\hspace*{5mm} Working in first quantization, the free-particle
Duffin\cite{duf}-Kemmer\cite{kem}-Pe\-tiau\cite{pet}
equation can be written as:
\begin{equation} \label{a}
\beta^{\mu} \partial_{\mu} \psi = m \psi
\end{equation}
where the $\beta^{\mu}$ are $16 \times 16$ complex matrices satisfying the
condition
\begin{equation}
\beta^{\mu} \beta^{\nu} \beta^{\lambda} + \beta^{\lambda} \beta^{\nu}
\beta^{\mu} = - g^{\mu \nu} \beta^{\lambda} - g^{\lambda \nu} \beta^{\mu} .
\end{equation}
This algebra splits into five (spin 0), ten (spin 1) and one-dimensional
(trivial) represen\-ta\-tions\cite{gen}.

Historically, the loss of interest in the DKP stems from the equivalence of
the DKP equation to the KG and the Proca equations\cite{fis,kra}, in
addition to the greater algebraic complexity of the DKP formulation.
We show how this equation leads to interesting results on the quaternion
field.\\ Before doing it we must remember that a
basic theorem\cite{bir} in the foundation of quantum mechanics states
that a general quantum mechanical system can be represented as a vector
space with scalar coefficients drawn from the real, the complex, the
quaternion\cite{fil,adl1,adl2} or the octonion\cite{gur} fields
(the scalar coefficients form a division algebra).
We assume a non commutative (but associative) multiplication in this paper,
and so our analysis applies to the quaternion field. We show (in
the simplest case - spin 0) the non equivalence between the {\em minimal}
DKP (spin 0) and the KG equation. In fact, we find that the KG
equation is equivalent to a \underline{pair} of DKP equations.

Quaternions are a generalization of the complex numbers
\begin{equation}
q =  z_{1} + j z_{2}
\end{equation}
\begin{center}
$(z_{m}\in {\cal C}(1,i) \hspace*{5mm} m = 1,2)$
\end{center}
with
\begin{equation}
i^{2} = j^{2} = k^{2} = -1 \: ; \hspace*{10mm} i j k = -1 \: .
\end{equation}

In our formalism the momentum operator is defined as
\begin{equation}
p^{\mu} = \partial^{\mu} \vert i
\end{equation}
where a {\em bared} operator $A\vert b$ acts as follows upon the
quaternion column matrix $\phi$
\begin{equation}
(A\vert b) \phi \: \equiv \: A \phi b \; \: .
\end{equation}
We anticipate that the only $b$ term appearing in this formalism is
$i$ (together, of course, with the trivial identity). We emphasize that a
characteristic of this formalism is the absolute need of a complex scalar
product ($CSP$), defined in terms of the quaternion counterpart ($QSP$) by
\begin{equation}
CSP = \frac{1 - i\vert i}{2} \; QSP
\end{equation}

The $CSP$ first introduced in 1984 by Horwitz and Biedenharn\cite{hor}, in
order to define consistently multiparticle quaternion states, has been
then justified by papers on Dirac equation\cite{rot}, representations of
$U(1,q)$\cite{del1} and translations between Quaternion and Complex Quantum
Mechanics\cite{del2}.\\ However, we remember that a different approach
is formulated by Adler who uses the $QSP$ (his fundamental results
are quoted in the recent book of ref.\cite{adl1}) and by Morita\cite{mor}
who uses complex quaternions ( or {\em biquaternions}) which contain an
additional commuting imaginary ${\cal I} = \sqrt{-1}$.

Let us now examine the simplest possible
case (spin 0) for the $DKP$ equation.\\
We start with the second order KG equation:
\begin{equation}
( \partial_{\mu} \partial^{\mu} + m^{2} ) \varphi = 0
\end{equation}
which can be rewritten into the form of a first-order matrix
differential equation
\[ \beta^{\mu} \partial_{\mu} \psi = m \psi \]
with
\begin{eqnarray}
\psi & = & \left( \begin{array}{c} \frac{\partial_{t} \varphi}{m}\\
\frac{\partial_{x} \varphi}{m}\\ \frac{\partial_{y} \varphi}{m}\\
\frac{\partial_{z} \varphi}{m}\\
\varphi \end{array} \right)
\end{eqnarray}
and
\begin{eqnarray}
\begin{array}{ccccccc}
\beta^{0} & = & \left( \begin{array}{ccccc}
\cdot & \cdot & \cdot & \cdot & 1\\
\cdot & \cdot & \cdot & \cdot & \cdot\\
\cdot & \cdot & \cdot & \cdot & \cdot\\
\cdot & \cdot & \cdot & \cdot & \cdot\\
-1 & \cdot & \cdot & \cdot & \cdot
\end{array} \right) & , &
\beta^{1} & = & \left( \begin{array}{ccccc}
\cdot & \cdot & \cdot & \cdot & \cdot\\
\cdot & \cdot & \cdot & \cdot & 1\\
\cdot & \cdot & \cdot & \cdot & \cdot\\
\cdot & \cdot & \cdot & \cdot & \cdot\\
\cdot & 1 & \cdot & \cdot & \cdot
\end{array} \right)\\ \\
\beta^{2} & = & \left( \begin{array}{ccccc}
\cdot & \cdot & \cdot & \cdot & \cdot\\
\cdot & \cdot & \cdot & \cdot & \cdot\\
\cdot & \cdot & \cdot & \cdot & 1\\
\cdot & \cdot & \cdot & \cdot & \cdot\\
\cdot & \cdot & 1 & \cdot & \cdot
\end{array} \right) & , &
\beta^{3} & = & \left( \begin{array}{ccccc}
\cdot & \cdot & \cdot & \cdot & \cdot\\
\cdot & \cdot & \cdot & \cdot & \cdot\\
\cdot & \cdot & \cdot & \cdot & \cdot\\
\cdot & \cdot & \cdot & \cdot & 1\\
\cdot & \cdot & \cdot & 1 & \cdot
\end{array} \right)
\end{array}
\end{eqnarray}

By returning to the $CSP$ we
must note that $1$ and $j$ are two orthogonal solutions,
therefore we have the appearance of all four standard Dirac free-particle
solutions\cite{rot} notwithstanding the two-component structure of the
wave function
and two orthogonal solutions, corresponding to spin up and spin down, in the
Schr\"odinger equation ({\em belated theoretical discovery of spin}).\\
However we also find two solutions (for a given four-moment) for the KG
equation with the result that in addition to the normal scalar, we discover
an {\em anomalous} scalar\cite{del3}. Such doubling of solutions is also in
the DKP equation but contrary to the KG equation, it is a matrix equation
and we can therefore hope to reduce it on the quaternion field.

In the next Section we shall find the quaternion matrix which reduces
the $5 \times 5$ complex matrix into two $3 \times 3$ quaternion matrix
blocks. In the subsequent Section we study  the pair
of quaternion $DKP$ equations and prove their equivalence.
In Section IV we explicitly find
the {\em new} KG equations corresponding to the pair of quaternion
DKP equations. The physical significance of these results and the possible
applications of the `quaternion' relativistic equation with {\em anomalous}
solution in quantum field theory
are discussed in our conclusions in the final Section.

\section{A {\em particular} quaternion reduction of the
$5 \times 5$ complex $\beta^{\mu}$ matrices}

\hspace*{5 mm} In our precedent paper\cite{del2} we have defined a set of
rules for translating from standard Quantum Mechanics ($QM$) to a particular
version of Qua\-ter\-nion Quan\-tum Me\-cha\-nics ($QQM$).
We must emphasize that this translation is limited to {\em even} complex
matrices. How can we translate the $5 \times 5$ complex matrices of the DKP
equation now?\\
We prove in this section with a {\em particular} reduction on the
quaternion field that the $5 \times 5$ complex $\beta^{\mu}$ matrices split
into two ``overlapping'' $3 \times 3$ quaternion matrix blocks. We have in this
{\em particular} reduction the following relation between
complex [$cmd$] and quaternion [$qmd$] matrix dimensions:
\begin{equation} \label{b}
(2n + 1) \; [cmd] \; \; \rightarrow  \; \; 2 \; [ overlapping \; \; blocks]
\; \times \; (n+1) \; [qmd] \; .
\end{equation}
To demonstrate the relation (\ref{b}) we start with the standard DKP
equation and analyse the simplest case (spin 0). We thus work with
$5 \times 5$ complex matrices. Since we use a complex geometry ($CSP$) we
have normal ($\sim a+ib$) and anomalous  ($\sim j(a+ib)$) solutions.\\
We shall find the explicit form of the matrix necessary to reduce the
complex matrices on the quaternion field. Not to disturb our next
con\-si\-derations by a complicate mathematical lan\-guage let us indicate with
\begin{eqnarray} \label{c}
\begin{array}{ccccccccc}
\left( \begin{array}{c} 1\\ \cdot\\ \cdot\\ \cdot\\
\cdot \end{array} \right) & , &
\left( \begin{array}{c} \cdot\\ 1\\ \cdot\\ \cdot\\
\cdot \end{array} \right) & , &
\left( \begin{array}{c} \cdot\\ \cdot\\ 1\\ \cdot\\
\cdot \end{array} \right) & , &
\left( \begin{array}{c} \cdot\\ \cdot\\ \cdot\\ 1\\
\cdot \end{array} \right) & , &
\left( \begin{array}{c} \cdot\\ \cdot\\ \cdot\\ \cdot\\
1 \end{array} \right)
\end{array}
\end{eqnarray}
the normal solutions and with
\begin{eqnarray} \label{d}
\begin{array}{ccccccccc}
\left( \begin{array}{c} j\\ \cdot\\ \cdot\\ \cdot\\
\cdot \end{array} \right) & , &
\left( \begin{array}{c} \cdot\\ j\\ \cdot\\ \cdot\\
\cdot \end{array} \right) & , &
\left( \begin{array}{c} \cdot\\ \cdot\\ j\\ \cdot\\
\cdot \end{array} \right) & , &
\left( \begin{array}{c} \cdot\\ \cdot\\ \cdot\\ j\\
\cdot \end{array} \right) & , &
\left( \begin{array}{c} \cdot\\ \cdot\\ \cdot\\ \cdot\\
j \end{array} \right)
\end{array}
\end{eqnarray}
the anomalous solutions.

In the next Section we shall explicitly find the solutions of the DKP
equation (spin 0). We shall have only two normal and two anomalous
solutions but for the considerations of this Section we can neglect it. By
returning to normal and anomalous solutions (\ref{c}), (\ref{d}) we
transform the first ones in the following column matrices
\begin{eqnarray} \label{e}
\begin{array}{ccccccccc}
\left( \begin{array}{c} 1\\ \cdot\\ \cdot\\ \cdot\\
\cdot \end{array} \right) & , &
\left( \begin{array}{c} j\\ \cdot\\ \cdot\\ \cdot\\
\cdot \end{array} \right) & , &
\left( \begin{array}{c} \cdot\\ 1\\ \cdot\\ \cdot\\
\cdot \end{array} \right) & , &
\left( \begin{array}{c} \cdot\\ j\\ \cdot\\ \cdot\\
\cdot \end{array} \right) & , &
\left( \begin{array}{c} \cdot\\ \cdot\\ 1\\ \cdot\\
\cdot \end{array} \right)
\end{array}
\end{eqnarray}
and the second ones in
\begin{eqnarray} \label{f}
\begin{array}{ccccccccc}
\left( \begin{array}{c} \cdot\\ \cdot\\ \cdot\\ 1\\
\cdot \end{array} \right) & , &
\left( \begin{array}{c} \cdot\\ \cdot\\ \cdot\\ j\\
\cdot \end{array} \right) & , &
\left( \begin{array}{c} \cdot\\ \cdot\\ \cdot\\ \cdot\\
1 \end{array} \right) & , &
\left( \begin{array}{c} \cdot\\ \cdot\\ \cdot\\ \cdot\\
j \end{array} \right) & , &
\left( \begin{array}{c} \cdot\\ \cdot\\ j\\ \cdot\\
\cdot \end{array} \right)
\end{array}
\end{eqnarray}
by the matrix
\begin{eqnarray} \label{smat}
\begin{array}{ccc}
S & = & \left( \begin{array}{ccccc}
a & ja & \cdot & \cdot & \cdot\\
\cdot & \cdot & a & ja & \cdot\\
\cdot & \cdot & \cdot & \cdot & 1\\
-jd & d & \cdot & \cdot & \cdot\\
\cdot & \cdot & -jd & d & \cdot
\end{array} \right)
\end{array}
\end{eqnarray}
where
\[ a = \frac{1-i\vert i}{2} \; \; \; , \; \; \;
d = \frac{1+i\vert i}{2} \]
are {\em generalized\/}\cite{del2} quaternions.\\
We have effected this transformation with the hope to find two DKP equa\-tion,
one corresponding to the normal solutions (now representing by (\ref{e})
and the other corresponding to the anomalous solutions (now representing by
(\ref{f})).

By using the matrix~(\ref{smat}) we can rewrite the DKP equation (\ref{a})
in the fol\-lo\-wing way
\begin{equation}
(S \beta^{\mu} S^{+} \partial_{\mu} - m ) S \psi = 0 \; \; ,
\end{equation}
where
\begin{eqnarray*}
\begin{array}{ccc}
S^{+} & = & \left( \begin{array}{ccccc}
a & \cdot & \cdot & ja & \cdot\\
-jd & \cdot & \cdot & d & \cdot\\
\cdot & a & \cdot & \cdot & ja\\
\cdot & -jd & \cdot & \cdot & d\\
\cdot & \cdot & 1 & \cdot & \cdot
\end{array} \right) \; \; ,
\end{array}
\end{eqnarray*}
\[ (ja)^{+}=(j-ji\mid i)^{+}=(j+k\mid i)^{+}=-j+k\mid i=-jd \; \; , \; \;
a^{+}=a \; \; ,  \; \; d^{+}=d \; \; , \]
and
\[ S^{+} S = 1 \; \; .\]
The similarity transformation is therefore
\[ S \beta^{\mu}_{ocm} S^{+} = \beta^{\mu}_{nqm} \]
where
\begin{center}
$ocm$ = old complex matrices\\
$nqm$ = new quaternion matrices
\end{center}
We are ready to give explicitly the new quaternion matrices obtained by
the $S$ matrix (this reduction is for us very important if we want kill the
anomalous solution).\\
The new quaternion matrices are
\begin{eqnarray}
\begin{array}{ccc}
\beta^{0}_{nqm} = \left( \begin{array}{ccccc}
\cdot & \cdot & a & 0 & 0\\
\cdot & \cdot & \cdot & 0 & 0\\
-a & \cdot & \cdot & -ja & \cdot\\
0 & 0 & -jd & \cdot & \cdot\\
0 & 0 & \cdot & \cdot & \cdot
\end{array} \right) & , &
\beta^{1}_{nqm} =  \left( \begin{array}{ccccc}
\cdot & \cdot & ja & 0 & 0\\
\cdot & \cdot & \cdot & 0 & 0\\
-jd & \cdot & \cdot & d & \cdot\\
0 & 0 & d & \cdot & \cdot\\
0 & 0 & \cdot & \cdot & \cdot
\end{array} \right)\\ \\
\beta^{2}_{nqm} = \left( \begin{array}{ccccc}
\cdot & \cdot & \cdot & 0 & 0\\
\cdot & \cdot & a & 0 & 0\\
\cdot & a & \cdot & \cdot & ja\\
0 & 0 & \cdot & \cdot & \cdot\\
0 & 0 & -jd & \cdot & \cdot
\end{array} \right) & , &
\beta^{3}_{nqm} = \left( \begin{array}{ccccc}
\cdot & \cdot & \cdot & 0 & 0\\
\cdot & \cdot & ja & 0 & 0\\
\cdot & -jd & \cdot & \cdot & d\\
0 & 0 & \cdot & \cdot & \cdot\\
0 & 0 & d & \cdot & \cdot
\end{array} \right)
\end{array}
\end{eqnarray}
which we can split into
\begin{eqnarray} \label{g}
\begin{array}{ccccccc}
\beta^{0} & = & \left( \begin{array}{ccc}
\cdot & \cdot & a\\
\cdot & \cdot & \cdot\\
-a & \cdot & \cdot
\end{array} \right) & , &
\beta^{1} & = & j \; \left( \begin{array}{ccc}
\cdot & \cdot & a\\
\cdot & \cdot & \cdot\\
-d & \cdot & \cdot
\end{array} \right)\\ \\
\beta^{2} & = & \left( \begin{array}{ccc}
\cdot & \cdot & \cdot\\
\cdot & \cdot & a\\
\cdot & a & \cdot
\end{array} \right) & , &
\beta^{3} & = & j \; \left( \begin{array}{ccc}
\cdot & \cdot & \cdot\\
\cdot & \cdot & a\\
\cdot & -d & \cdot
\end{array} \right)
\end{array}
\end{eqnarray}
and
\begin{eqnarray} \label{h}
\begin{array}{ccccccc}
\tilde{\beta}^{0} & = & -j \; \left( \begin{array}{ccc}
\cdot & a & \cdot\\
d & \cdot & \cdot\\
\cdot & \cdot & \cdot
\end{array} \right) & , &
\tilde{\beta}^{1} & = & \left( \begin{array}{ccc}
\cdot & d & \cdot\\
d & \cdot & \cdot\\
\cdot & \cdot & \cdot
\end{array} \right)\\ \\
\tilde{\beta}^{2} & = & j \; \left( \begin{array}{ccc}
\cdot & \cdot & a\\
\cdot & \cdot & \cdot\\
-d & \cdot & \cdot
\end{array} \right) & , &
\tilde{\beta}^{3} & = & \left( \begin{array}{ccc}
\cdot & \cdot & d\\
\cdot & \cdot & \cdot\\
d & \cdot & \cdot
\end{array} \right)
\end{array}
\end{eqnarray}
Now the $\beta^{\mu}$ matrices act on the normal solutions and the
$\tilde{\beta}^{\mu}$ matrices on the anomalous solutions. Note that
$a=\frac{1-i\vert i}{2}$, therefore the $\beta^{\mu}$ matrices (\ref{g})
act on the third element of the column matrix by cancelling the $j, k$
part. The form of the solution of the DKP equation with $\beta^{\mu}$
matrices is like
\begin{eqnarray}
\left( \begin{array}{c} z_{1} +j z_{2}\\
z_{3} +j z_{4}\\ z_{5} \end{array} \right)
\end{eqnarray}
with
\[ z_{m} \in {\cal C} \; \; m=1,2,...,5 \; . \]
In the same way (now we must note that $d=\frac{1+i\vert i}{2}$) we can
prove that the form of the solution of the DKP equation with
$\tilde{\beta}^{\mu}$
matrices (\ref{h}) is like
\begin{eqnarray}
\left( \begin{array}{c} j \tilde{z}_{5}\\
\tilde{z}_{1} +j \tilde{z}_{2}\\
\tilde{z}_{3} +j \tilde{z}_{4} \end{array} \right)
\end{eqnarray}
with
\[ \tilde{z}_{m} \in {\cal C} \; \; m=1,2,...,5 \; . \]
We shall prove in the next Section that the solutions to the DKP equations
can be written as
\begin{center}
{\em normal DKP equation}
\end{center}
\begin{eqnarray} \label{v}
\psi_{normal} & = & \left( \begin{array}{c}
\frac{(\partial_{t} + j \partial_{x})\varphi}{m}\\
\frac{(\partial_{y} + j \partial_{z})\varphi}{m}\\
\varphi \end{array} \right)
\end{eqnarray}
\begin{center}
{\em anomalous DKP equation}
\end{center}
\begin{eqnarray} \label{y}
\psi_{anomalous} & = & \left( \begin{array}{c}
j \varphi\\
\frac{(\partial_{t} + j \partial_{x})\varphi}{m}\\
\frac{(\partial_{y} + j \partial_{z})\varphi}{m}
\end{array} \right)
\end{eqnarray}
with $\varphi$ complex wave functions.

\section{The quaternion DKP equation}

\hspace*{5 mm} We analyse the DKP equation (\ref{a}) with the
${\beta}^{\mu}$ matrices given by (\ref{g}). We can now proceed in the
standard manner. The plane wave solutions are of the form (n.b. the
ordering)
\[ \psi (\vec{x},t) = u (\vec{p} \; ) e^{-ipx} \]
with the `spinors' $u$ satisfies the relation
\begin{equation} \label{i}
(\beta^{\mu} p_{\mu} \vert i + m ) u = 0 \; .
\end{equation}
To simplify our consideration we start with $\vec{p} = 0 $ and find
the solutions to the eq.(\ref{i}). We have
\[ - {\beta}^{0} p_{0} u i = m u \; . \]
Explicitly
\begin{eqnarray}
\begin{array}{ccc}
ap_{0} \left( \begin{array}{c}
-u_{3}\\ 0\\ u_{1} \end{array} \right) i & = &
m  \left( \begin{array}{c}
u_{1}\\ u_{2}\\ u_{3} \end{array} \right) \; .
\end{array}
\end{eqnarray}
Therefore if $p_{0} = + m$ the plane wave solution is
\begin{eqnarray} \label{p}
\frac{1}{\sqrt{2}} \left( \begin{array}{c}
-i\\ 0\\ 1 \end{array} \right) \; e^{-imt}
\end{eqnarray}
whereas if $p_{0} = - m$ is
\begin{eqnarray} \label{w}
\frac{1}{\sqrt{2}} \left( \begin{array}{c}
+i\\ 0\\ 1 \end{array} \right) \; e^{+imt} \; .
\end{eqnarray}
Note that the normalization condition is now:
\[ - \psi^{+} \eta \beta^{0} \psi i \]
with
\[ \eta = - (2 (\beta^{0})^{2} + 1) \]
(in fact we must remember that, for the DKP equation, the continuity
equa\-tion is
\[-\partial_{\mu} ( \psi^{+} \eta \beta^{\mu} \psi i) = 0 \; ) \; \; . \]
Now we are ready to find the DKP solutions for $\vec{p}\neq 0$. By
remembering the eq.(\ref{i}) and the explicit representation of
$\beta^{\mu}$ matrices (\ref{g}) we have
\begin{equation}
\left( \begin{array}{c}
-p_{0}au_{3}i+p_{x}jau_{3}i\\
p_{y}au_{3}i+p_{z}jau_{3}i\\
p_{0}au_{1}i-p_{x}jdu_{1}i+p_{y}au_{2}i-p_{z}jdu_{1}i
\end{array} \right) = m \; \left( \begin{array}{c}
u_{1}\\ u_{2}\\ u_{3} \end{array} \right)
\end{equation}
and therefore
\begin{eqnarray}
\begin{array}{ccc}
u_{1} & = & -\frac{ip_{0}+kp_{x}}{m} (u_{3})_{\cal C}\\ \\
u_{2} & = & \frac{ip_{y}-kp_{z}}{m} (u_{3})_{\cal C}\\ \\
u_{3} & = & (u_{3})_{\cal C}
\end{array}
\end{eqnarray}
with $\cal C$ we indicate the complex projection
\[ (q)_{\cal C} = \frac{1-i\vert i}{2} q \; . \]
To find, in the limit $\vec{p} \rightarrow 0$, the solutions
(\ref{p}, \ref{w}) we pose $(u_{3})_{\cal C}=\frac{1}{\sqrt{2}}$
\begin{eqnarray} \label{z}
\psi_{(p_{0}=+\vert p_{0}\vert )} & = & \frac{1}{\sqrt{2}}
\left( \begin{array}{c}
-\frac{i\vert p_{0}\vert +kp_{x}}{m}\\
\frac{ip_{y}-kp_{z}}{m}\\
1 \end{array} \right) \; e^{-ipx}
\end{eqnarray}
\begin{eqnarray} \label{x}
\psi_{(p_{0}=-\vert p_{0}\vert )} & = & \frac{1}{\sqrt{2}}
\left( \begin{array}{c}
\frac{i\vert p_{0}\vert -kp_{x}}{m}\\
\frac{ip_{y}-kp_{z}}{m}\\
1 \end{array} \right) \; e^{-ipx} \; .
\end{eqnarray}
The normal solution are now represented by (\ref{z}, \ref{x}). In the same
way we can prove that the anomalous solutions are represented by
\begin{eqnarray}
{\psi}_{(p_{0}=+\vert p_{0}\vert)} & = & \frac{1}{\sqrt{2}}
\left( \begin{array}{c}
j\\
-\frac{i\vert p_{0}\vert+kp_{x}}{m}\\
\frac{ip_{y}-kp_{z}}{m} \end{array} \right) \; e^{-ipx}
\end{eqnarray}
\begin{eqnarray}
{\psi}_{(p_{0}=-\vert p_{0}\vert)} & = & \frac{1}{\sqrt{2}}
\left( \begin{array}{c}
j\\
\frac{i\vert p_{0}\vert-kp_{x}}{m}\\
\frac{ip_{y}-kp_{z}}{m} \end{array} \right) \; e^{-ipx} \; .
\end{eqnarray}

We conclude this Section by noting that the representations of the
ma\-tri\-ces
(\ref{g}, \ref{h}) are equivalent. In fact one can prove that the following
matrix
\begin{eqnarray}
S & = & \left( \begin{array}{ccc}  \cdot & \cdot & j\\
1 & \cdot & \cdot\\ \cdot & 1 & \cdot  \end{array} \right)
\end{eqnarray}
transforms the $\beta^{\mu}$ in the $\tilde{\beta}^{\mu}$ matrices and
consequently we can connect the normal to the anomalous solutions by
\begin{equation}
{\psi}_{anomalous} = S {\psi}_{normal}
\end{equation}
With help of the DKP equation we can kill the anomalous so\-lu\-tion.
The\-re\-fo\-re we find, with quaternions, an important result:
\begin{center}
``{\em the $KG$ equation is equivalent to a pair of $DKP$ equations}''.
\end{center}

\section{The {\em new} KG equation on the quaternion field}

\hspace*{5mm} Our purpose, in this Section, is to show that with
quaternions we have the possibility to choose either the standard or the
{\em new} ( {\em modified}) $KG$ equation.\\
Before doing it, we must find
the {\em new} $KG$
equation on the quaternion field. We start with the equation (\ref{a})
(with matrices (\ref{h}) and  anomalous solution (\ref{y})).
If we explicitly write this equation we have
\begin{equation}
(-ja\partial_{t}+d\partial_{x})\frac{(\partial_{t} +
j \partial_{x})\varphi}{m} + (ja\partial_{y}+d\partial_{z})\frac{(\partial_{y}
+ j \partial_{z})\varphi}{m} = m (j\varphi)
\end{equation}
By noting that $d$ kills the complex wave function $\varphi$ and that $ja=
dj$ we can rewrite the previous equation in the following way
\begin{equation}
ja(\partial_{\mu}\partial^{\mu} + m^{2}) \varphi =
d(\partial_{\mu}\partial^{\mu} + m^{2}) \tilde{\varphi} = 0
\end{equation}
with
\begin{center}
$\tilde{\varphi}=j\varphi$ {\em pure} quaternion function.
\end{center}
It is simple to adapt it to $DKP$
equation with matrices (\ref{g}) and normal solution (\ref{v}) finding
\begin{equation}
a (\partial_{\mu}\partial^{\mu} + m^{2}) \varphi = 0
\end{equation}
These equation can be related by a similarity transformation, in fact:
\begin{equation}
-j \; \frac{1-i\vert i}{2} \; j \; \; = \; \; \frac{1+i\vert i}{2}
\end{equation}
and, consequently, we can go from standard to anomalous solution by $j$
multiplication.

The situation in $QQM$ (Quaternion Quantum Mechanics) can be summa\-ri\-ze as
follows:
\begin{center}
{\em Standard Klein-Gordon equation (four plane wave solutions)}
\end{center}
\begin{equation} \label{j}
\alpha (\partial_{\mu}\partial^{\mu} + m^{2}) \phi = 0 \; \; ,
\end{equation}
\begin{center}
($\alpha = 1$) ,\\
\begin{tabular}{ll}
$e^{-ipx}$ & positive and negative energy ,\\
$je^{-ipx}$ & complex-orthogonal solutions ;
\end{tabular}
\end{center}
\begin{center}
{\em New Klein-Gordon equation (two plane wave solutions)}
\end{center}
\begin{equation} \label{k}
\alpha (\partial_{\mu}\partial^{\mu} + m^{2}) \phi = 0
\end{equation}
\begin{center}
($\alpha = \frac{1-i\vert i}{2}$ complex solutions $e^{-ipx}$ ,\\
 $\alpha = \frac{1+i\vert i}{2}$ {\em pure} quaternion solutions
$je^{-ipx}$) ,
\end{center}
obviously $\phi$ is a quaternion function.

{}From now on we use the {\em new} Klein-Gordon equation when we want kill
the anomalous solutions and the standard K-G equation when we want
consider normal and anomalous solution. It is important to note that in our
formalism the $DKP$ equation is replaced by the {\em new} K-G equation.

\section{Conclusions}

\hspace*{5mm} In this work we have studied the Duffin-Kemmer-Petiau
equation from the point of view of Quaternion Quantum Mechanics with a
complex geometry. In Complex Quantum Mechanics the Klein-Gordon equation is
equivalent to the {\em spin $0$} part of the Duffin-Kemmer-Petiau equation
(which presents a greater algebraic complexity). In our formalism the DKP
equation (which kil\-ls the anomalous solutions) is {\em not} equivalent to
the ``stan\-dard'' KG equation
($\partial_{\mu}\partial^{\mu}\phi+m^{2}\phi=0$).\\
Therefore, for the first time in Quaternion Quantum Mechanics, the
{\em new} (or {\em modified}) Klein-Gordon equation appears.

Our consideration have been limited to the {\em spin $0$ part} of the Duffin-
Kemmer-Petiau equation but it is very simple to extend them to the
{\em spin $1$ part} of the Duffin-Kemmer-Petiau equation (there the
standard and {\em new} Proca equations should appear).

Our goal has been to demonstrate the equivalence between the
{\em quaternion} Duffin-Kem\-mer-Petiau equation and the {\em new} Klein-
Gordon equation.

The situation in the Quaternion Quantum Mechanics can be sum\-ma\-ri\-ze as
fol\-lows:
\begin{center}
\begin{tabular}{|l|l|}     \hline
standard Dirac equation & contains \\
- $4\times4$ complex $\gamma^{\mu}$ matrices - &
the anomalous solutions \\ \hline
{\em quaternion} Dirac equation & kills\\
- see ref.~\cite{rot} - & the anomalous solutions \\ \hline \hline
standard Klein-Gordon equation & contains\\
- see eq.~(\ref{j}) - & the anomalous solutions \\ \hline
{\em new} Klein-Gordon equation & kills\\
- see eq.~(\ref{k}) - & the anomalous solutions. \\  \hline
\end{tabular}
\end{center}

Nevertheless we are {\em not} obliged to kill the anomalous solutions. The
correct equations and the corresponding Lagrangian is in practice determined
only when the number of particles in the theory is fixed. For example in
the Higgs sector of the Electroweak theory\cite{sal,wei,gla} we used in our
recent paper\cite{del4} the standard Klein-Gordon equation which
automatically contains for us four particles corresponding to the four
Higgs ${\cal H}^{0}$, ${\cal H}^{+}$, ${\overline{\cal H}}^{0}$,
${\cal H}^{-}$ before spontaneous symmetry breaking. We shall also
{\em resuscitate} the standard Dirac equation in the fermion sector of the
Electroweak theory where we interpret the normal solution as the
{\em neutrino}-field and the anomalous solution as the {\em electron}-field.
The operator which distinguish between normal and anomalous solutions in
the {\em Quaternion Electroweak Theory} ({\em QET}) is repre\-sen\-ted by the
third
component ($i\vert i$) of weak-isospin\cite{del4}. A complete and detailed
{\em QET} will be presented in our forthcoming paper.

Before this work we had the possibility of choice between equations with and
without anomalous solutions only for the fermion equation whereas for the
bosonic equation we were always obliged to consider equation {\em with}
anomalous solutions. Now we have finally the possibility to work in
Qua\-ter\-nion Quantum Mechanics using bosonic equation {\em without\/}
anomalous so\-lu\-tions. Nevertheless we remember (also once)
that the equations with anomalous so\-lu\-tions (very important in the
{\em Quaternion Electroweak Theory}) \underline{don't disappear}.

We hope that this paper points out the non triviality in the choice to
adopt quaternions as the underlying number field.

\end{document}